\documentclass[aps,prl,superscriptaddress,showpacs,floatfix,twocolumn]{revtex4}
\usepackage{graphicx}   % Include figure files

\begin{document}

\title{Suppression of High-$p_\mathrm{T}$ Neutral Pion Production 
in Central Pb+Pb Collisions at $\sqrt{s_\mathrm{NN}} = 17.3$~GeV
Relative to p+C and p+Pb Collisions}

%%%****************************  New Feature ******************************
%%%%%%%%%%%%%%%%%%%%%%%%%%%%%%%%%%%%%%%%%%%%%%%%%%%% PRL length check lines

\author{ 
 M.M.~Aggarwal,$^{1}$
 Z.~Ahammed,$^{2}$
 A.L.S.~Angelis,$^{3, *}$
 V.~Antonenko,$^{4}$
 V.~Arefiev,$^{5}$
 V.~Astakhov,$^{5}$
 V.~Avdeitchikov,$^{5}$
 T.C.~Awes,$^{6}$
 P.V.K.S.~Baba,$^{7}$
 S.K.~Badyal,$^{7}$
% ...
} 

\medskip
\affiliation{(WA98 Collaboration)}
%\author
\medskip

\affiliation{$^{1}$~University of Panjab, Chandigarh 160014, India}
\affiliation{$^{2}$~Variable Energy Cyclotron Centre, Calcutta
   700064, India}
\affiliation{$^{3}$~University of Geneva, CH-1211 Geneva
   4,Switzerland}
\affiliation{$^{4}$~RRC ``Kurchatov Institute'',
   RU-123182 Moscow}
\affiliation{$^{5}$~Joint Institute for Nuclear Research,
   RU-141980 Dubna, Russia}
\affiliation{$^{6}$~Oak Ridge National
   Laboratory, Oak Ridge, Tennessee 37831-6372, USA}
\affiliation{$^{7}$~University of Jammu, Jammu 180001, India}
\affiliation{$^{8}$~University of M{\"u}nster, D-48149 M{\"u}nster,
   Germany}
\affiliation{$^{9}$~SUBATECH, Ecole des Mines, Nantes, France}
\affiliation{$^{10}$~University of Rajasthan, Jaipur 302004, Rajasthan,
   India}
\affiliation{$^{11}$~University of Lund, SE-221 00 Lund, Sweden}
\affiliation{$^{12}$~MIT Cambridge, MA 02139}
\affiliation{$^{13}$~Institute of Physics, Bhubaneswar 751005,
   India}
\affiliation{$^{14}$~University of Tsukuba, Ibaraki 305, Japan}
\affiliation{$^{15}$~Universiteit
   Utrecht/NIKHEF, NL-3508 TA Utrecht, The Netherlands}
\affiliation{$^{16}$~KVI, University of Groningen, NL-9747 AA Groningen,
   The Netherlands} 
\affiliation{$^{17}$~Gesellschaft f{\"u}r Schwerionenforschung (GSI),
   D-64220 Darmstadt, Germany}
\affiliation{$^{18}$~Nuclear Physics Institute, CZ-250 68 Rez, Czech Rep.}
\affiliation{$^{19}$~Institute for Nuclear Studies,
   00-681 Warsaw, Poland}
\affiliation{$^{20}$~University of Tennessee, Knoxville,
   Tennessee 37966, USA}
\affiliation{$^{*}$Deceased}

%
%%%%%%%%%%%%%%%%%%%%%%%%%%%%%%%%%%%%%%%%%%%%%%%%%%%% end of length check lines

\date{\today}

\begin{abstract}
  Neutral pion transverse momentum spectra were measured in p+C and
  p+Pb collisions at $\sqrt{s_\mathrm{NN}} = 17.4$~GeV at mid-rapidity
  ($2.3 \lesssim \eta_\mathrm{lab} \lesssim 3.0$) over the range $0.7
  \lesssim p_\mathrm{T} \lesssim 3.5$~GeV/$c$. The spectra are
  compared to $\pi^0$ spectra measured in Pb+Pb collisions at
  $\sqrt{s_\mathrm{NN}} = 17.3$~GeV in the same experiment. For a wide
  range of Pb+Pb centralities ($N_\mathrm{part} \lesssim 300$)
  the yield of $\pi^0$'s with $p_\mathrm{T} \gtrsim 2$~GeV/$c$ is
  larger than or consistent with the p+C or p+Pb yields scaled with the
  number of nucleon-nucleon collisions ($N_\mathrm{coll}$), while for
  central Pb+Pb collisions with $N_\mathrm{part} \gtrsim 350$
  the $\pi^0$ yield is suppressed.
\end{abstract}

    % For heavy ion papers we usually use:
    % \pacs{25.75.Dw}

% It is optional to also add (uncomment):
% \keywords{}

\maketitle

The study of hadron production at high transverse momentum
($p_\mathrm{T}$) is a sensitive tool to characterize the matter
created in ultrarelativistic heavy-ion collisions, and in particular,
to detect the possible formation of a quark-gluon plasma (QGP), i.e.,
a thermalized phase in which quarks and gluons are the relevant
degrees of freedom.  Particles at high $p_\mathrm{T}$ result from
quark and gluon scatterings with high momentum transfer (``hard
scattering'') which can be described by perturbative
quantum-chromodynamics (pQCD). The scattered quarks and gluons will
traverse the created medium and fragment into the observable hadrons.
High-$p_\mathrm{T}$ particle production in nucleus-nucleus (A+A)
collisions was predicted to be suppressed~\cite{Wang:1991xy} as a
consequence of the energy loss of the scattered partons in the dense
matter (``jet quenching'').  Such suppression was observed by
experiments at the Relativistic Heavy Ion Collider (RHIC) in central
Au+Au and Cu+Cu collisions at a center-of-mass energy of up to
$\sqrt{s_\mathrm{NN}} = 200$~GeV \cite{Adcox:2004mh,Adams:2005dq,
  Arsene:2004fa,Phobos_whitepaper,Klein-Boesing:2006kd}. Within
jet-quenching models the suppression
% of up to a factor 5 in central Au+Au collisions 
can be related to medium properties, such as the initial gluon density
or the transport coefficient
\cite{Vitev:2002pf,Vitev:2005he,Kovner:2003zj}.

A crucial test for the idea of parton energy loss in the hot and dense
medium created in A+A collisions is the measurement of the
$\sqrt{s_\mathrm{NN}}$ dependence of high-$p_\mathrm{T}$ hadron
production
\cite{Wang:1998hs,Wang:1998ww,Zhang:2001ce,Vitev:2002pf,Vitev:2005he,
  Adare:2008cx}.  In central Pb+Pb collisions at the CERN SPS energy
of $\sqrt{s_\mathrm{NN}} = 17.3$~GeV ($p_\mathrm{lab} = 158~A$GeV/$c$)
the initial energy density, as estimated from the measured transverse
energy~\cite{Aggarwal:2000bc}, is above the critical value
$\varepsilon_\mathrm{c} \approx 1$~GeV/fm$^3$~\cite{Cheng:2006qk} for
the transition to the QGP.  On the other hand, the initial gluon
density and the lifetime of a deconfined phase produced at
$\sqrt{s_\mathrm{NN}} = 17.3$~GeV will be significantly reduced as
compared to RHIC energies.  Results at SPS energies thereby provide a
sensitive test of jet quenching model predictions.

Results on high-$p_\mathrm{T}$ particle production in central Pb+Pb
collisions at the CERN SPS have already been
published~\cite{Aggarwal:2001gn,Blume:2006va}.  However, the
interpretation of these data has been complicated by the lack of
reference p+p data to allow to quantify nuclear effects.  Instead,
parameterizations of p+p data have been employed to search for nuclear
effects \cite{Aggarwal:2001gn,d'Enterria:2004ig}, but with substantial
systematic uncertainties.  Moreover, hadron suppression due to parton
energy loss might be compensated by an enhancement due to multiple
soft scatterings of the incoming partons prior to the hard scattering
process (``nuclear $k_\mathrm{T}$-enhancement'' or ``Cronin effect'').
Measurements in p+A and d+A collisions at different
$\sqrt{s_\mathrm{NN}}$ suggest that such enhancement is significantly
stronger at $\sqrt{s_\mathrm{NN}} \approx 20$~GeV than at
$\sqrt{s_\mathrm{NN}} = 200$~GeV \cite{Wang:1998ww}.

In this letter $\pi^0$ spectra are presented from p+C and p+Pb
collisions at $\sqrt{s_\mathrm{NN}} = 17.4$~GeV ($p_\mathrm{lab} =
160$~GeV/$c$) measured in the WA98 experiment.  Since the nuclear
$k_\mathrm{T}$-enhancement in p+C is expected to be small
\cite{Barnafoldi:2003kb} this measurement should provide a useful
substitute for a p+p reference.  Information on the magnitude of the
nuclear $k_\mathrm{T}$-enhancement may be obtained by comparison of
the p+Pb and p+C spectra.  The spectra are compared to $\pi^0$ spectra
measured in Pb+Pb collisions at $\sqrt{s_\mathrm{NN}} = 17.3$~GeV with
the same WA98 experimental setup~\cite{Aggarwal:2001gn}.

In the WA98 experiment $\pi^0$ yields were measured by detection of
photons from the $\pi^0 \rightarrow \gamma\gamma$ decay branch with a
highly-segmented lead-glass calorimeter. This detector was located
21.5\,m downstream from the target and subtended the pseudorapidity
range $2.3 \lesssim \eta_\mathrm{lab} \lesssim 3.0$.  A 400 GeV/$c$
proton beam from the CERN SPS impinged on a beryllium production
target to provide a mixed secondary beam selected to have momentum of
160~GeV/c.  The secondary beam consisted primarily of protons and
pions with roughly equal content. Protons were identified with use of
two gas Cherenkov counters located upstream of the 1879 mg/cm$^2$
$^{12}$C (495 mg/cm$^2$ $^{208}$Pb) target. The WA98 minimum bias
trigger condition required a minimum amount of transverse energy
($E_\mathrm{T}$) in the region $3.5 \lesssim \eta_\mathrm{lab}
\lesssim 5.5$, measured with a sampling calorimeter with
electromagnetic and hadronic sections.  The measured minimum bias
cross section $\sigma_\mathrm{mb}$ for p+C (p+Pb) of 193\,mb
(1422\,mb) corresponds to 86\,\% (81\,\%) of the total geometric cross
section.  The number of analyzed minimum bias events was $1.2 \cdot
10^6$ ($1.0 \cdot 10^6$) for p+C (p+Pb).  A high-energy photon (HEP)
trigger based on the sum energy signal of overlapping $4\times 4$
groups of towers in the lead-glass calorimeter was used to enhance the
sample of high $p_\mathrm{T}$ events.  The efficiency of the HEP
trigger reached $\sim 100$\,\% for photons with $p_\mathrm{T} \gtrsim
0.8$~GeV/$c$.  An additional $1.5 \cdot 10^6$ ($0.5 \cdot 10^6$) p+C
(p+Pb) HEP events were analyzed which corresponded to $3.9 \cdot 10^7$
($8.2 \cdot 10^6$) sampled minimum bias events.

Neutral pion yields were determined statistically by counting photon
pairs with invariant mass in the $\pi^0$ mass range after subtraction
of the normalized background from uncorrelated pairs.  The shape of
this background was determined by mixing photons from different
events. Only photon pairs with an energy asymmetry $\alpha = |E_1 -
E_2|/(E_1 + E_2) < 0.7$ were used in the analysis. A correction for
geometrical acceptance and reconstruction efficiency was applied to
the raw $\pi^0$ yields.  The reconstruction efficiency takes into
account the loss of $\pi^0$'s due to the photon identification and
energy asymmetry cuts.  It also modifies the $\pi^0$ yield to account
for the $p_\mathrm{T}$ shift that results from the finite energy
resolution of the lead-glass calorimeter convoluted with the steeply
falling $\pi^0$ $p_\mathrm{T}$ spectrum.  Effects of overlapping
showers in the calorimeter, which were important in central Pb+Pb
collisions, are negligible in p+C and p+Pb collisions.  The dominant
systematic uncertainties are listed in Table~\ref{tab:syserr}.  The
systematic uncertainties of the peak extraction, acceptance
correction, and efficiency correction are approximately independent of
$p_\mathrm{T}$.  The energy scale of the calorimeter was confirmed by
comparison of the measured $p_\mathrm{T}$-dependent $\pi^0$ peak
positions with GEANT simulations. The estimated uncertainty of
$1.5\,\%$ on the energy scale leads to an uncertainty on the $\pi^0$
yields that increases with $p_\mathrm{T}$.

\begin{table}[h]
\caption{\label{tab:syserr}
Systematic uncertainties (in $\%$) on the $\pi^0$ yields
in p+C and p+Pb collisions at three $p_\mathrm{T}$ values.
}
\begin{ruledtabular}
\begin{tabular}{lccc}
$p_\mathrm{T}$ (GeV/$c$)           &  1        & 2        & 3    \\
\hline
peak extraction                    &  6        & 6        & 6    \\

geometric acceptance               &  2.5      & 2.5      & 2.5  \\

$\pi^0$ reconstruction efficiency  &  11       & 11       & 11   \\

energy scale                       &  5        & 10       & 20   \\

\hline
total                              &  14       & 16       & 24   \\

\end{tabular}
\end{ruledtabular}
\end{table}

%%%%%%%%%%%%%%%%%%%%%%%%%%%%%%%%%%%%%%%% Figure 1.
\begin{figure}
\includegraphics[width=1.0\linewidth]{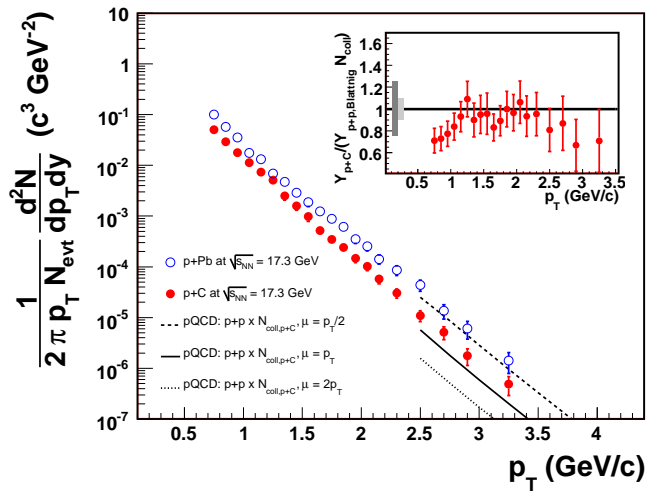}
\caption{\label{fig:spectra} Invariant $\pi^0$ yields in minimum bias
  p+C and p+Pb collisions at $\sqrt{s_\mathrm{NN}} = 17.4$~GeV. The
  error bars represent the quadratic sum of statistical and systematic
  uncertainties. The lines represent next-to-leading-order pQCD
  calculation of the $\pi^0$ yield in $\sqrt{s} = 17.4$~GeV p+p
  collisions for three different scales, scaled by $\langle
  N_\mathrm{coll}^\mathrm{p+C} \rangle = 1.7$. The inset shows a
  comparison of the p+C spectrum to a parameterization of the yield in
  p+p collisions at $\sqrt{s} = 17.3$~GeV from \cite{Blattnig:2000zf}.
  The dark gray box indicates the uncertainty of the parameterization
  estimated in \cite{d'Enterria:2004ig}, the light gray box the
  uncertainty of $\langle N_\mathrm{coll}^\mathrm{p+C} \rangle$.}
\end{figure}

The spectra of the invariant $\pi^0$ yields in p+C and p+Pb collisions
at $\sqrt{s_\mathrm{NN}} = 17.4$~GeV are shown in
Fig.~\ref{fig:spectra}.  The transition between the minimum bias and
the HEP sample occurs at $p_\mathrm{T} = 1.7$~GeV/$c$. The p+C
spectrum is compared to a next-to-leading-order (NLO) pQCD calculation
for p+p at the same energy \cite{deFlorian:2005yj}. The calculation
was performed with the CTEQ6M \cite{Pumplin:2002vw} parton
distribution functions and the ``Kniehl-Kramer-P{\"o}tter'' (KKP) set
of fragmentation functions \cite{Kniehl:2000hk} with renormalization
and factorization scales set to be the same at $\mu = p_\mathrm{T}/2$,
$p_\mathrm{T}$, or $2 p_\mathrm{T}$.  For comparison with the p+C data
the pQCD calculations have been scaled by the average number of
nucleon-nucleon collisions in p+C.  The pQCD calculation shows a large
uncertainty related to the arbitrary choice of scale. It has been
suggested that the NLO perturbative expansion is not sufficient at low
energies and that threshold resummation corrections must be taken into
account~\cite{deFlorian:2005yj}.  The large theoretical uncertainties
demonstrate that pQCD calculations cannot be used as a reliable
reference at CERN SPS energies.  The inset of Fig.~\ref{fig:spectra}
shows that the $\pi^0$ yields per nucleon-nucleon collision measured
in p+C are in good agreement with a parameterization of $\pi^0$
spectra in p+p from Blattnig et al.~\cite{Blattnig:2000zf} that has
been employed to study nuclear effects in Pb+Pb
collisions~\cite{d'Enterria:2004ig}.

Nuclear effects in $\pi^0$ production can be quantified with the
nuclear modification factor defined as
\begin{equation}
R'_\mathrm{AA} = \frac{\langle N_\mathrm{coll}^\mathrm{p+B} \rangle}
                     {\langle N_\mathrm{coll}^\mathrm{A+A} \rangle} 
  \frac{\left. \mathrm{d}N_{\pi^0}/\mathrm{d}p_\mathrm{T}\right|_\mathrm{A+A}}
       {\left. \mathrm{d}N_{\pi^0}/\mathrm{d}p_\mathrm{T}\right|_\mathrm{p+B}} 
  \;.
\label{eq:raa}
\end{equation}
In the absence of nuclear effects $R'_\mathrm{AA}$ is expected to be
unity for $p_\mathrm{T} \gtrsim 2$\,GeV/$c$ where hard scattering is
expected to dominate particle production. $\langle N_\mathrm{coll}
\rangle$ was determined with a Glauber Monte Carlo calculation
\cite{Miller:2007ri} using an inelastic nucleon-nucleon cross section
of $\sigma_\mathrm{inel}^\mathrm{NN} = (31.8 \pm 2)$\,mb
\cite{Alt:2005zq}.  The same Glauber calculation was used to extract
$\langle N_\mathrm{coll} \rangle$ values in Pb+Pb collisions.  In the
Glauber calculation the transverse energy $E_\mathrm{T}$ was modelled
by sampling a negative binomial distribution to determine the
$E_\mathrm{T}$ contribution of each participating nucleon
\cite{Miller:2007ri}.  The bias due to the trigger selection on the
measured $E_\mathrm{T}$ was taken into account. For minimum bias p+C
and p+Pb collisions values of $\langle N_\mathrm{coll}
\rangle_\mathrm{p+C} = 1.7 \pm 0.2$ and $\langle N_\mathrm{coll}
\rangle_\mathrm{p+Pb} = 3.8 \pm 0.4$ were obtained.

The $\langle N_\mathrm{coll} \rangle$ values for Pb+Pb collisions were
determined by applying cuts to the simulated $E_\mathrm{T}$ that
corresponded to the same fraction of
$\sigma_\mathrm{mb}^\mathrm{Pb+Pb}$ as the cuts applied to the
measured $E_\mathrm{T}$. These $\langle N_\mathrm{coll} \rangle$
values are listed in Table~\ref{tab:ncoll} and agree within systematic
uncertainties with those of Ref.~\cite{Aggarwal:2001gn} that were
determined with the VENUS 4.12 event generator in which
$\sigma_\mathrm{inel}^\mathrm{NN} = 29.6$\,mb was used.  With the
large acceptance of the WA98 $E_\mathrm{T}$ measurement, and good
description of the $E_\mathrm{T}$ distribution, including
fluctuations~\cite{Aggarwal:2000bc}, a centrality class corresponding
to the $1\,\%$ most central Pb+Pb collisions could be defined to
access very large $\langle N_\mathrm{coll} \rangle$ values.

\begin{table}[h]
  \caption{\label{tab:ncoll}
    Results of the Glauber calculation for Pb+Pb collisions at 
    $\sqrt{s_\mathrm{NN}} = 17.3$~GeV ($\sigma_\mathrm{inel}^\mathrm{NN} = 31.8$\,mb). 
    Centrality classes are given as a fraction (\%) of 
    $\sigma_\mathrm{mb}^\mathrm{Pb+Pb} \approx 6300$~mb \cite{Aggarwal:2001gn}.
    The systematic uncertainty of $\langle N_\mathrm{coll} \rangle$  
    was parameterized as 
    $\sigma^2_{N_\mathrm{coll}} = 0.1^2 + (0.4 \exp(-N_\mathrm{coll}/100))^2$. 
  }
\begin{ruledtabular}
\begin{tabular}{lccccccccc}
class     
   &  1   & 2    & 3    & 4     & 5     & 6     & 7     & 8     & $6-8$   \\
from 
   &  82.8& 67.0  & 48.8 & 25.3  & 13.0  & 6.8   & 1.0  & 0     & 0 \\
to   
   &  100 & 82.8  & 67.0 & 48.8  & 25.3  & 13.0  & 6.8  & 1.0   & 13.0 \\
$\langle N_\mathrm{part} \rangle$ 
   & 8.2  & 23   & 54.2 & 123.2 & 218.2 & 289.1 & 347.8 & 382.6 & 322.5 \\
$\langle N_\mathrm{coll} \rangle$ 
   & 6.3  & 22.1 & 67.1 & 202.9 & 433.1 & 627.0 & 803.7 & 912.0 & 727.8 \\
\end{tabular}
\end{ruledtabular}
\end{table}

\begin{figure}
\includegraphics[width=1.0\linewidth]{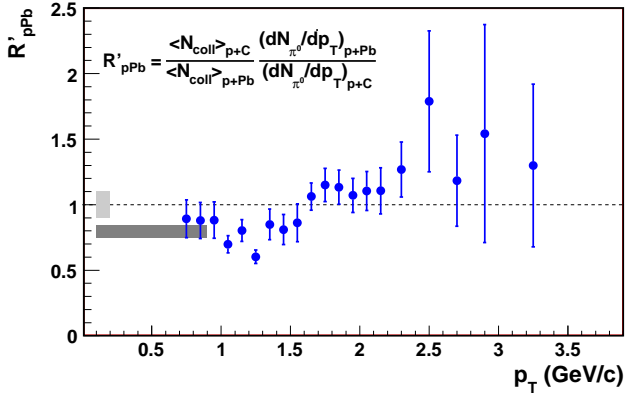}
\caption{\label{fig:pb_c_ratio} The $N_\mathrm{coll}$-normalized ratio
  of $\pi^0$ yields in p+Pb and p+C collisions. The box at
  $R'_\mathrm{pPb} \approx 0.8$ represents the expectation for the
  case of scaling with $N_\mathrm{part}$ rather than
  $N_\mathrm{coll}$. The box at unity indicates the $N_\mathrm{coll}$
  systematic uncertainty. The error bars represent the quadratic sum 
  of the statistical and remaining systematic uncertainties. The energy 
  scale uncertainty was assumed to cancel. 
}
\end{figure}

The p+Pb $\pi^0$ spectrum appears to be flatter than the p+C spectrum
(see Fig.~\ref{fig:spectra}).  The ratio of the
$N_\mathrm{coll}$-normalized p+Pb and p+C spectra is shown in
Fig.~\ref{fig:pb_c_ratio}.  At low $p_\mathrm{T}$ ($\sim 1$~GeV/$c$)
the ratio is consistent with scaling with the number of participating
nucleons ($N_\mathrm{part} = N_\mathrm{coll} + 1$ in p+A) while
scaling with $N_\mathrm{coll}$ is observed at higher $p_\mathrm{T}$
($\sim 2$~GeV/$c$). In fact, for $p_\mathrm{T} \gtrsim 2$~GeV/$c$ the
ratio tends to be above unity, consistent with an expected stronger
nuclear $k_\mathrm{T}$-enhancement in p+Pb than in p+C.
% as expected in most models.

%%%%%%%%%%%%%%%%%%%%%%%%%%%%%%%%%%%%%%%% Figure 2.

%%%%%%%%%%%%%%%%%%%%%%%%%%%%%%%%%%%%%%%% Figure 3.
\begin{figure}
\includegraphics[width=1.0\linewidth]{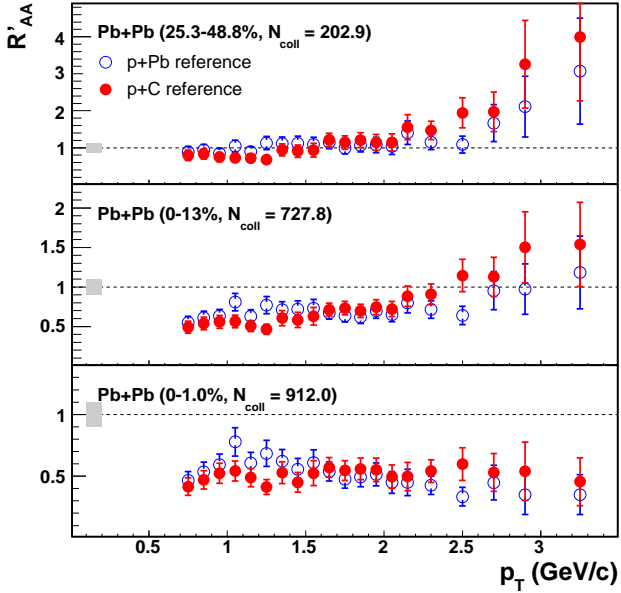}
\caption{\label{fig:raa} $\pi^0$ $R'_\mathrm{AA}$ in Pb+Pb collisions
  at $\sqrt{s_\mathrm{NN}} = 17.3$~GeV for three centrality classes
  using p+C or p+Pb spectra as a reference. The boxes around unity
  reflect the systematic uncertainties related to $\langle
  N_\mathrm{coll} \rangle$.}
\end{figure}

Neutral pion spectra from Pb+Pb collisions at $\sqrt{s_\mathrm{NN}} =
17.3$~GeV were published by WA98 in \cite{Aggarwal:2001gn}.
Fig.~\ref{fig:raa} shows the $\pi^0$ nuclear modification factor
$R'_\mathrm{AA}$ as defined in Eq.~\ref{eq:raa} for three Pb+Pb
collision centralities. For the $25.3-48.8\,\%$ (of
$\sigma_\mathrm{mb}^\mathrm{Pb+Pb}$) most central collisions the data
points at high $p_\mathrm{T}$ ($p_\mathrm{T} \gtrsim 2$~GeV/$c$)
suggest a stronger nuclear $k_\mathrm{T}$-enhancement than in p+C.  On
the other hand, for the $0-13\,\%$ most central Pb+Pb collisions
$R'_\mathrm{AA}$ is smaller than for the $25.3-48.8\,\%$ class but
still consistent with unity at high $p_\mathrm{T}$. For the $0-1\,\%$
most central collisions the $\pi^0$ yield is significantly suppressed
compared to either p+Pb or p+C as reference.  An apparent suppression
of the $\pi^0$ yield in very central Pb+Pb collisions compared to
peripheral collisions was noted in Ref.~\cite{Aggarwal:2001gn}.
However the large uncertainty of using the peripheral distribution as
a reference did not allow to draw a firm conclusion.  The observed
suppression is qualitatively consistent with expectations from
jet-quenching.

\begin{figure}
\includegraphics[width=1.0\linewidth]{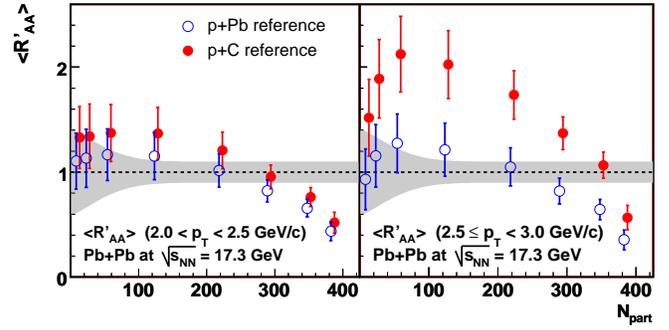}
\caption{\label{fig:raa_vs_npart} Centrality dependence of the average
  $\langle R'_\mathrm{AA} \rangle$ for $2 < p_\mathrm{T} <
  2.5$~GeV/$c$ and $2.5 \leq p_\mathrm{T} < 3.0$~GeV/$c$. The bands
  around unity indicate the systematic uncertainty of $\langle
  N_\mathrm{coll} \rangle$. The error bars represent the quadratic sum
  of the statistical and systematic uncertainties. The
  $N_\mathrm{part}$ values for the full points were shifted by +5 for
  better visibility.}
\end{figure}

The centrality dependence of the average $R'_\mathrm{AA}$ for $2 <
p_\mathrm{T} < 2.5$~GeV/$c$ and $2.5 \leq p_\mathrm{T} < 3.0$~GeV/$c$ is
shown in Fig.~\ref{fig:raa_vs_npart}. The $\pi^0$ yields in Pb+Pb in
these $p_\mathrm{T}$ ranges are not suppressed for $N_\mathrm{part}
\lesssim 300$ ($\langle R'_\mathrm{AA} \rangle \gtrsim 1$).  For
more central Pb+Pb collisions $\langle R'_\mathrm{AA} \rangle$
decreases with centrality indicating significant suppression of the
high $p_\mathrm{T}$ $\pi^0$ yield.  The apparent lack of suppression,
or enhancement even, for $N_\mathrm{part} \lesssim 300$ may be
due to competing effects of suppression due to parton energy loss and
nuclear $k_\mathrm{T}$-enhancement.

%%%%%%%%%%%%%%%%%%%%%%%%%%%%%%%%%%%%%%%% Figure 4.

In summary, $\pi^0$ spectra were measured in minimum bias p+C and p+Pb
collisions at $\sqrt{s_\mathrm{NN}} = 17.4$~GeV in the range $0.7
\lesssim p_\mathrm{T} \lesssim 3.5$~GeV/$c$. Based on these spectra
the nuclear modification factors $R'_\mathrm{AA}$ for Pb+Pb collisions
at CERN SPS energies could be determined using a measured p+A
reference. In very central Pb+Pb collisions ($0-1\,\%$ of
$\sigma_\mathrm{mb}^\mathrm{Pb+Pb}$) a significant suppression of
high-$p_\mathrm{T}$ neutral pions was observed ($R'_\mathrm{AA}
\approx 0.5 \pm 0.14$) that is reminiscent of the high-$p_\mathrm{T}$
hadron suppression observed in Cu+Cu and Au+Au collisions at RHIC. The
pion suppression reported here, together with the results at higher
energies from RHIC, constrain jet-quenching models and may help
clarify the behavior of the energy loss of partons in strongly
interacting matter.

%%%%%%%%%%%%%%%%%%%%%%%%%%%%%%%%%%%%%%%%%%%%%%%%%%%%%%%%%%%%
% Acknowledgements
%
\begin{acknowledgments}
  We would like to thank the CERN-SPS accelerator crew for the
  excellent beam provided and the Laboratoire National Saturne for the
  loan of the magnet Goliath.  We thank W. Vogelsang for providing the
  QCD calculation shown in this paper. This work was supported jointly
  by the German BMBF, DFG, and the Helmholtz-Gemeinschaft (VI-VH-146),
  the U.S.  DOE, the Swedish NFR, the Dutch Stichting FOM, the Swiss
  National Fund, the Humboldt Foundation, the Stiftung f\"{u}r
  deutsch-polnische Zusammenarbeit, the Department of Atomic Energy,
  the Department of Science and Technology and the University Grants
  Commission of the Government of India, the Indo-FRG Exchange
  Programme, the PPE division of CERN, the INTAS under contract
  INTAS-97-0158, the Polish KBN under the grant 2P03B16815, the
  Grant-in-Aid for Scientific Research (Specially Promoted Research \&
  International Scientific Research) of the Ministry of Education,
  Science, Sports and Culture, JSPS Research Fellowships for Young
  Scientists, the University of Tsukuba Special Research Projects, and
  ORISE.  ORNL is managed by UT-Battele, LLC, for the U.S. Department
  of Energy under contract DE-AC05-00OR22725.
\end{acknowledgments}

%\section{Acknowledgements}

%REFERENCES:  Use \begin{references} and \end{references}.  Do not use
%             \begin{thebibliography} and \end{thebibliography}.
%             You may either
%       (a) enter all citations explicitly or
%               (b) use some "\def" shorthand notations.
%             Our first paper used approach (a) and our second used (b).
%             Here are the two reference lists as examples of how to proceed:

\bibliography{WA98-25_9}

%FIGURES:  Place all the figures here (after the references) in sequence.

%
% We strongly recommend that you create the reference section using
% BibTeX:
%       \bibliography{basename of .bib file}
% This will ensure that you get most of the references exactly right,
% because they can be downloaded from SPIRES in bibtex format:
% http://131.169.91.193/spires/hep/ or from .bib files used in previous
% PHENIX or other publications.  Futhermore, using BibTeX will ensure
% that you get all references cited in numerical order, because BibTeX
% takes care of that for you.
%
% If you insist to go it alone, you may explicitly enter references:

% The last uncommented line must be:

\end{document}